\documentclass{article}
\usepackage{spconf}
\usepackage{cite}
\usepackage{url}
\usepackage{textcomp}

\usepackage{multirow}

\usepackage{amsmath,amssymb,amsfonts}
\usepackage{stfloats}
\usepackage{nicefrac}

\usepackage{graphicx}
\usepackage[caption=false, font=footnotesize]{subfig}
\usepackage{xcolor}

\usepackage[acronym]{glossaries}

\usepackage{algorithm}

\usepackage{amsmath}
\usepackage{mathtools}

\usepackage{enumitem}
\usepackage{array}
\usepackage{amsthm}
\usepackage[titletoc,toc,title]{appendix}
\usepackage[bookmarksopen=true]{hyperref}
\usepackage[capitalise,nameinlink]{cleveref}

\usepackage{nccmath}
\usepackage{bookmark}

\usepackage{csquotes}
\usepackage{url}
\usepackage{mathrsfs}

\usepackage{algorithmic}
\usepackage{scrextend}

\graphicspath{Figures/}
\newcommand{\tr}{^\top}			
\newcommand{\bs}[1]{\boldsymbol{#1}}	
\allowdisplaybreaks[4]					
\long\def\comment#1{}

\newfont{\bbb}{msbm10 scaled 700}

\newcommand{\cv}{{\bf c}}

\newcommand{\xv}{{\bf x}}
\newcommand{\yv}{{\bf y}}

\newcommand{\Fm}{{\bf F}}
\newcommand{\Gm}{{\bf G}}

\newcommand{\Id}{{\bf I}}

\newcommand{\Pm}{{\bf P}}

\newcommand{\Rm}{{\bf R}}
\newcommand{\Sm}{{\bf S}}
\newcommand{\Tm}{{\bf T}}
\newcommand{\Um}{{\bf U}}

\newcommand{\Vm}{{\bf V}}
\newcommand{\Xm}{{\bf X}}
\newcommand{\Ym}{{\bf Y}}

\newcommand{\Oc}{{\cal O}}

\renewcommand{\arg}{{\hbox{arg}}}

\theoremstyle{remark}

\newcommand{\figsub}[2]{%
  \hyperref[#1]{\cref*{#1}(#2)}%
}

\title{FaSST: Fast Sparsifying Secondary Transform} 
\name{\hspace{-2.75em}Darukeesan Pakiyarajah$^{\star}$, Samuel Fernández-Menduiña$^{\star}$, Eduardo Pavez$^{\star}$, Antonio Ortega$^{\star}$, {Debargha Mukherjee$^{\dagger}$} \begin{NoHyper}
\thanks{This work was supported in part by a grant from Google.}
\end{NoHyper}}
\address{$^{\star}$University of Southern California, Los Angeles, CA, USA \\
$^{\dagger}$Google LLC, Mountain View, CA, USA}

\begin{document}
\ninept
\maketitle
\begin{abstract}
Data-dependent secondary transforms, which aim to decorrelate coefficients of a separable primary transform, can improve residual coding efficiency; however, their deployment is often constrained by computational complexity. Recent video codecs use variants of the low-frequency non-separable transform (LFNST), which discards some high-frequency secondary transform coefficients, limiting achievable coding gains. Moreover, existing data-dependent secondary transforms lack explicit rate–distortion (RD) optimal design criteria. In this work, we propose a framework for designing low-complexity data-dependent secondary transforms, termed Fast Sparsifying Secondary Transforms (FaSSTs). Our approach approximates data-driven sparse orthonormal transforms (SOTs) by factorizing them into a sequence of Givens rotations. The rotations are efficiently determined using an alternating minimization strategy combined with an approximate Givens factorization procedure.  Our method adapts the number of rotations based on the prediction mode, further reducing computational complexity.
We design mode-dependent secondary transforms for intra-prediction residuals in AV2 using FaSST. Experimental results show that mode-adaptive FaSST matches the RD performance of LFNST while reducing the number of computations by $83.67\%$. Moreover, by avoiding fixed-coefficient truncation, FaSST achieves up to $1.80\%$ BD-rate savings relative to LFNST while operating at $66.24\%$ lower complexity.
\end{abstract}
\begin{keywords}
data-dependent transforms, secondary transforms, fast transforms, Givens rotation, rate-distortion optimization
\end{keywords}
\section{Introduction}
\label{sec:intro}
\begin{figure}[t]
    \centering
    \includegraphics[width=1\linewidth]{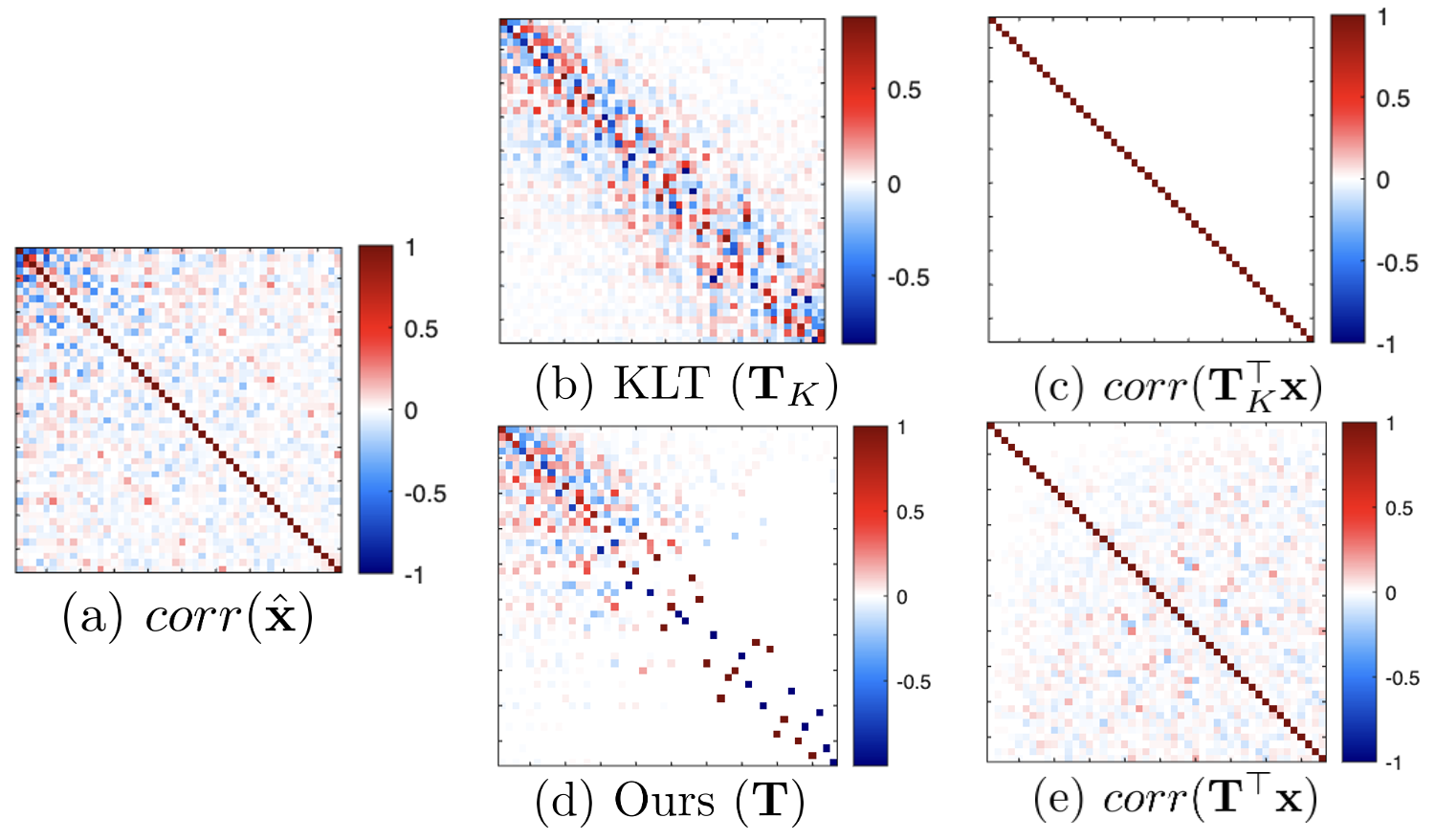}
    \vskip -3ex
    \caption{Correlation matrices and transform kernels for AV2 intra-prediction residuals ($D_{135}$ mode): (a) correlation of 48 (out of $16^2$) sorted DCT coefficients, (b) KLT kernel ($48 \times 48$), (c) correlation after KLT, (d) proposed FaSST ($48 \times 48$), and (e) correlation after the proposed transform.}
    \label{fig:correlation}
    \vskip -4ex
\end{figure}

The non-separable Karhunen–Loève Transform (KLT) is optimal for linear decorrelation under common statistical assumptions \cite{Zhu11,Yeo12MDDT, fan2019signal, xu2012video, arrufat2014non}. 
However, the non-separable KLT is rarely used in practice due to its lack of low-complexity implementations and memory requirements. Separable trigonometric transforms, such as the discrete cosine transform (DCT), are widely adopted in modern codecs because they approximate the KLT for certain types of signals with low complexity. However, separable transforms are less effective at decorrelating directional residuals. As shown in \figsub{fig:correlation}{a}, residual correlations remain in the transform coefficients, mainly among low-frequency components and adjacent higher frequencies. 
To account for these correlations under practical complexity constraints, transform coding in modern block-based codecs is often organized into two stages: a separable primary transform, applied directly to prediction residual blocks, and a non-separable secondary transform, applied to a subset of primary transform coefficients to further decorrelate them \cite{bross2021overview, Koo19}.

State-of-the-art methods design secondary transforms as KLTs computed from the low-frequency coefficients of the primary transform~\cite{bross2021overview,zhao2021study}. The  KLT kernel shown in \figsub{fig:correlation}{b} produces optimally decorrelated secondary transform coefficients, as illustrated in \figsub{fig:correlation}{c}. However, since the KLT kernel is dense,  perfect decorrelation of the selected primary transform coefficients comes at the cost of substantially higher computational complexity. In addition, the KLT design ignores quantization effects. As a result, coefficients that are likely to be quantized to zero are still decorrelated, leading to wasted computation. Moreover, because the KLT design does not account for sparsity or quantization noise, it yields suboptimal  performance at low bitrates \cite{Sezer15sot}. 
Although KLT approximations are used in practice, there remains interest in fast or structured RD-aware solutions for secondary transforms \cite{Koo19, Said16high}.

The most widely adopted approximations are variants of low-frequency non-separable transforms (LFNST) \cite{Koo19}, in which the KLT is applied to the selected low-frequency primary transform coefficients, and the higher-frequency secondary transform coefficients are not computed (i.e., forced to zero) to reduce computational cost.
However, fixed coefficient truncation prevents low-distortion reconstructions, limiting the coding gains from using secondary transforms in low-bitrate regimes. 
Another approach \cite{Said16high} factorizes KLT bases into sequences of low-cost Givens rotations, which are $2\times 2$ kernels that serve as key building blocks in low-complexity methods such as the fast Fourier transform (FFT) \cite{cooley1965algorithm}. 
In this mode-dependent approach, the secondary KLTs for each mode are approximated using Givens rotations, where the number of rotations controls the computational complexity. This method allocates equal numbers of rotations to all modes, failing to account for variation in residual statistics across prediction modes. 
For instance, horizontal and vertical modes are well aligned with separable primary transform directions, whereas diagonal modes such as DDR exhibit residual statistics that are harder to decorrelate with separable transforms. Accordingly, fewer rotations in the secondary transform are sufficient for horizontal and vertical modes as compared to DDR. 
This method ignores this property and searches for an RD–optimal fixed number of rotations across all modes using a brute-force strategy.  
%
%
%
All these techniques first design a KLT and then approximate it. However, sparse orthonormal transforms (SOTs) have been demonstrated to yield improved RD performance compared to KLT-based designs \cite{Sezer15sot,sezer2011robust}. SOTs are designed using an RD-inspired cost function that favors sparsity of transform coefficients and accounts for coefficients likely to be quantized to zero. Nonetheless, SOTs are dense kernels, which suffer from high computational complexity. 

In this work, to overcome the complexity and rate-distortion (RD) limitations of KLT-based secondary transforms, we propose a method to derive low-complexity approximations of SOTs using sequences of Givens rotations, yielding fast and structured secondary transforms referred to as Fast Sparsifying Secondary Transforms (FaSST). 
The main contributions of this work are threefold:
(i) we propose an optimization framework for deriving low-complexity approximations of SOTs using Givens rotations;
(ii) we empirically show that simpler structured transforms achieve better compression efficiency than aggressive coefficient truncation; 
(iii) we demonstrate that mode-adaptive allocation of computational budget yields substantial computational savings.

We reformulate the SOT problem by imposing a  constraint that expresses the transform as a product of Givens rotations. Our framework combines SOT’s alternating minimization strategy with an approximate Givens factorization procedure, enabling adaptive selection of the number of rotations while accounting for quantization effects, thereby achieving an improved complexity–gain trade-off. As shown in \figsub{fig:correlation}{d}, the proposed FaSST kernel is significantly sparser (approximately $55\%$ fewer non-zero entries) than the corresponding KLT in \figsub{fig:correlation}{b}, while still removing the most significant correlations (\figsub{fig:correlation}{e}).  
We empirically evaluate the proposed FaSST for designing mode-dependent secondary transforms using intra-prediction residual data from AV2 under a realistic coding setup. Results show that FaSST achieves RD performance comparable to LFNST while using $66.67\%$ fewer computations than LFNST with a fixed number of rotations. With mode-adaptive rotation placement, the computational cost can be further reduced by $83.67\%$ relative to LFNST while maintaining comparable RD performance. Furthermore, at a complexity level $66.24\%$ lower than that of LFNST, FaSST achieves up to a $1.80\%$ BD-rate saving.

\section{Preliminaries}
\subsection{Secondary transforms}
Let $\xv \in \mathbb{R}^{N^2}$ be the vector obtained by stacking the columns of an $N \times N$ block. Given a primary transform $\Gm$, and a permutation matrix $\Pm$ that reorders the transform coefficients (e.g., zig-zag scan), we write the primary transform coefficients as $\hat\xv = \Pm \Gm \tr \xv$. Let $\tilde\Tm$ be an $n \times n$ non-separable orthonormal transform, referred to as the \emph{secondary transform}, which is applied to the first $n$ low-frequency coefficients after reordering~\cite{Zhao16nsst,Koo19,Saxena12stip}. Let \(\mathbf T=\operatorname{blkdiag}(\tilde{\mathbf T},\mathbf I)\). Then
$\hat{\mathbf y} = ( \mathbf G \mathbf P \mathbf T )^\top \mathbf x$, with inverse transform $\xv = \Gm \Pm \Tm \hat\yv$. 

To reduce encoder complexity, state-of-the-art video codecs \cite{bross2021overview} introduce a series of simplifications on the pipeline above. First, they use a reduced KLT, where only the first $n_k$ low-frequency coefficients after the secondary transform are computed~\cite{bross2021overview,zhao2021study}. This reduces complexity from $\Oc(n^2)$ for full KLT to $\Oc(n_k n)$. Second, the high-frequency primary transform coefficients are zeroed out, resulting in $\mathbf T_r = \operatorname{blkdiag}(\tilde\Tm_{[:,\:1:n_k]}, \bf 0)$. As a result of this ``coefficient dropping'', 
$( \mathbf G \mathbf P \mathbf T_r )^\top$ is non-invertible, preventing low-distortion reconstructions. 
Hence, during RD optimization (RDO), the encoder is less likely to select secondary transforms at high bitrates.

\subsection{Sparse-orthonormal transforms}
\label{sec:sot}
Given training samples $\{\xv\}_{i=1}^{m_e}$, and a sparsity regularization parameter $\mu\geq 0$, the sparse-orthogonal transform (SOT) solves \cite{Sezer15sot}
\begin{equation}
\label{eq:sotprob}
\underset{\Fm,\{\hat{\yv}_i\}_{i=1}^{m_e}}{\text{min }} \;\; \sum_{i=1}^{m_e} \|{\xv}_i - \Fm \hat{\yv}_i\|_2^2 + \mu \|\hat{\yv}_i\|_0, \quad \text{s.t.} \;\; \Fm\tr\Fm=\Id,
\end{equation}
where $\|\cdot\|_0$ is the $\ell_0$ norm promoting sparsity of the transform coefficients $\hat{\yv}_i$ and $\Fm$ is the learned SOT. To account for quantization, $\mu$ is set to $\mu = (Q_s/2)^2$, where $Q_s$ denotes the quantization step size used in a target encoder. The squared-error term in \eqref{eq:sotprob} is a proxy for the distortion introduced by quantization, while the $\ell_0$ norm of the coefficients is a proxy for the coding rate. Hence, \eqref{eq:sotprob} can be interpreted as RD-inspired transform optimization.

A two-step alternating minimization algorithm~\cite{o1998alternating} is used to approximately solve \eqref{eq:sotprob}. The first step fixes the transform and finds:
\begin{equation}
    \label{eq:l0opt}
    \underset{\hat{\yv}_i}{\min \:\:}\|{\xv}_i - \Fm \hat{\yv}_i\|_2^2 + \mu \, \|\hat{\yv}_i\|_0,
\end{equation}
for $ i=1, \hdots, m_e$, which is solved via  hard thresholding~\cite{Sezer15sot, lorenz2008convergence}:
\begin{equation}
\label{eq:hth}
[\hat{\yv}^*_i]_j
= [\hat{\cv}_i]_j \, \mathbf{1}\!\left\{ |[\hat{\cv}_i]_j| \ge \mu^{1/2} \right\},
\end{equation}
where $\hat{\cv}_i = \Fm\tr \hat{\xv}_i$ and $\mathbf{1}\lbrace\cdot\rbrace$ is the indicator function. This step approximates the training samples using a small number of nonzero transform-domain coefficients. The second step optimizes the transform given the coefficients: 
\begin{align}
\label{eq:tropt}
    \underset{\Fm}{\text{min }} \;\;
& \sum_{i=1}^{m_e} \|{\xv}_i - \Fm \hat{\yv}_i\|_2^2 \qquad 
\text{s.t.} \;\;
\Fm\tr\Fm = \Id,
\end{align}
which has solution $\Fm^* = \Vm \Um\tr$ \cite{schonemann1966generalized}, where $\Um$ and $\Vm$ are the left and right singular vector matrices from the SVD of $\hat{\Ym}{\Xm}\tr$. Here, ${\Xm} \in \mathbb{R}^{n \times m_e}$ and $\hat{\Ym} \in \mathbb{R}^{n \times m_e}$ are the matrices formed by stacking $\{{\xv}_i\}_{i=1}^{m_e}$ and $\{\hat{\yv}_i\}_{i=1}^{m_e}$ from the previous step.


\section{Fast Sparsifying Secondary Transforms}
\label{sec:faSST}
LFNSTs significantly improve residual coding efficiency but rely on coefficient truncation to control computational complexity, which limits achievable coding gains. Moreover, KLTs and their approximations do not account for quantization effects. 
We address these limitations via our proposed Fast Sparsifying Secondary Transforms (FaSST). 
The following subsections present the main components of the proposed framework: the reformulation of the SOT problem under the Givens rotation constraint (\cref{sec:optimization}), the alternating minimization procedure used to solve it (\cref{sec:alt_min}), and the approximate Givens factorization method with mode-adaptive selection of the number of rotations (\cref{sec:givensrot}).
\subsection{Optimization problem}
\label{sec:optimization}
Hereinafter, we use $\hat{\xv}$ to denote the first $n$ low-frequency coefficients after the primary transform. Suppose we are given $m_e$ training examples, denoted by $\{\hat{\xv}_i\}_{i=1}^{m_e}$. We aim to learn a secondary transform parameterized as a sequence of Givens rotations. Modifying the problem in \eqref{eq:sotprob}, for $\mu \geq 0$, we seek
\begin{align}
\label{eq:givensprob}
\underset{\Sm_J,\{\hat{\yv}_i\}_{i=1}^{m_e}}{\text{min }} \;\;
& \sum_{i=1}^{m_e} \|\hat{\xv}_i - \Sm_J \hat{\yv}_i\|_2^2 + \mu \|\hat{\yv}_i\|_0 \\
\text{s.t.} \;\;
& \Sm_J = \prod_{j=1}^J \Gm(m_j, n_j, \theta_j), \notag
\end{align}
where $\Sm_J$ is the learned FaSST (secondary transform), the $\ell_0$ term promotes sparsity of the secondary transform coefficients $\hat{\yv}_i$, and $\Gm(m_j, n_j, \theta_j)$ is a Givens rotation acting on the
$(m_j,n_j)$ coordinates with angle $\theta_j$, i.e., given an input $\xv$, the output is
\[
\begin{aligned}
[\Gm(m_j,n_j,\theta_j)\xv]_{m_j} &= 
\cos(\theta_j)x_{m_j} + \sin(\theta_j)x_{n_j},\\
[\Gm(m_j,n_j,\theta_j)\xv]_{n_j} &=
-\sin(\theta_j)x_{m_j} + \cos(\theta_j)x_{n_j},
\end{aligned}
\]
while all other coordinates remain unchanged. The orthogonality constraint in \eqref{eq:sotprob} is no longer required,  since 
$\Sm_J$ is orthonormal by construction. In contrast to \eqref{eq:sotprob}, our formulation targets fast secondary transforms by explicitly seeking to determine the Givens rotations, i.e., $J$ and $\{(m_i,n_i,\theta_i) \}_{i=1}^J$. We next describe the alternating minimization procedure adopted to solve this problem.


\subsection{Alternating minimization steps}
\label{sec:alt_min}
We solve \eqref{eq:givensprob} using an alternating minimization strategy similar to that used for \eqref{eq:sotprob} \cite{Sezer15sot}, consisting of two steps that are iterated until the objective function converges: 1) for a fixed transform, we solve the inner minimization, which has the closed form solution in \eqref{eq:hth}, and 2) given the coefficients $\{\hat{\yv}_i\}_{i=1}^{m_e}$, we optimize the transform: 
\begin{align}
\label{eq:tropt2}
    \underset{\Sm_J}{\text{min }} \;\;
& \sum_{i=1}^{m_e} \|\hat{\xv}_i - \Sm_J \hat{\yv}_i\|_2^2 \qquad 
\text{s.t.} \;\;
\Sm_J = \prod_{j=1}^J \Gm(m_j, n_j, \theta_j).
\end{align}
It can be shown that the above problem is equivalent to 
\begin{align}
    \label{eq:trace_min}
    \underset{\Sm_J}{\max} \;\; &\mathrm{tr}(\bs\Gamma\Sm_J) &
\text{s.t.} \;\;
\Sm_J = \prod_{j=1}^J \Gm(m_j, n_j, \theta_j),
\end{align}
where $\mathrm{tr}(\cdot)$ denotes the trace of a matrix, $\bs{\Gamma} = \hat{\Ym}\hat{\Xm}\tr$ denotes the cross-covariance matrix between the thresholded secondary transform coefficients $\hat{\Ym}$ and the primary transform coefficients $\hat{\Xm}$,  and $\hat{\Xm} \in \mathbb{R}^{n \times m_e}$ and $\hat{\Ym} \in \mathbb{R}^{n \times m_e}$ are the matrices formed by stacking $\{\hat{\xv}_i\}_{i=1}^{m_e}$ and $\{\hat{\yv}_i\}_{i=1}^{m_e}$, respectively. We solve this problem using an iterative approach, in which Givens rotations are placed progressively by identifying the optimal rotation location $(m_i, n_i)$ and the rotation angle $\theta_i$, using an approximate Givens factorization procedure. The FaSST learning algorithm is summarized in \cref{alg:rdogr_algo}.




\begin{algorithm}[t]
\begin{algorithmic}[1]
\caption{FaSST Learning} \label{alg:rdogr_algo}
\renewcommand{\algorithmicrequire}{\textbf{Input:}}
 \renewcommand{\algorithmicensure}{\textbf{Output:}}
 \REQUIRE $\{\hat{\xv}_i\}_{i=1}^{m_e}, \tau, J_{\max}, \mu$
 \renewcommand{\algorithmicrequire}{\textbf{Initialize:}}
 \REQUIRE $\Sm_J$ is initialized to SOT
            \REPEAT
                \FOR{$i=1$ to $m_e$}
                \STATE $\hat{\yv}^*_i = \arg \: \underset{\hat{\yv}_i}{\min \:\:}\|\hat{\xv}_i - \Sm_J \hat{\yv}_i\|_2^2 + \mu \|\hat{\yv}_i\|_0$ \label{alg_step:st1}
                \ENDFOR
                \STATE $\bs\Gamma = \hat{\Ym}^*\hat{\Xm}$
                \STATE  $\{\Sm_J,J\} \gets $ Approximate Givens factorization $(\bs\Gamma, \tau, J_{\max})$\label{alg_step:st2}
            \UNTIL{convergence}
\ENSURE  $\Sm_J$ and $J$
\end{algorithmic}
\end{algorithm}

\begin{algorithm}[t]
\begin{algorithmic}[1]
\caption{Approximate Givens factorization} \label{alg:Jacobisvd}
\renewcommand{\algorithmicrequire}{\textbf{Input:}}
 \renewcommand{\algorithmicensure}{\textbf{Output:}}
 \REQUIRE $\bs\Gamma, \tau, J_{\max}$
 \renewcommand{\algorithmicrequire}{\textbf{Initialize:}}
 \REQUIRE $\Um_0=\Vm_0=\Sm_0=\Id$, $j = 1$, $e_0 = \inf$
            \WHILE {$e_j > \tau$ and $j < J_{\max}$}
            \STATE $\bs\Gamma_j \gets \Um_{j-1}\tr\bs\Gamma\Vm_{j-1}$
            \STATE $(m_j, n_j) = \arg\:\underset{\substack{(p,q), p>q \\ (p,q) \neq (m_i,n_i) \: \forall i<j}}{\max} \: |[\bs\Gamma_j\tr\bs\Gamma_j]_{pq}|$ \label{alg_step:st3}
            \STATE Compute $\alpha_j$ and $\beta_j$ in \eqref{eq:svds} using SVD.
            \STATE  $\Um_{j} \gets \Um_{j-1}\Gm(m_j,n_j,\alpha_j)$ 
            \STATE $\Vm_{j}\gets \Vm_{j-1}\Gm(m_j,n_j,\beta_j)$
            \STATE $\Sm_j \gets \Vm_j\Um_j\tr$
            \STATE Compute $e_j$ using \eqref{eq:diag_error}.
            \STATE $j \gets j+1$
            \ENDWHILE
            \STATE $J \gets j$ and $\Sm_J \gets \Sm_j$
\ENSURE  $\Sm_J$, $J$
\end{algorithmic}
\end{algorithm}

\subsection{Approximate Givens factorization procedure}
\label{sec:givensrot}
To derive an iterative algorithm to solve \eqref{eq:trace_min}, let $\Sm_J = \Vm_J \Um_J\tr$, where 
\begin{align}
    \label{eq:ujvj}
    \Um_J = \prod_{j=1}^J \Gm(m_j, n_j, \alpha_j), \qquad  \Vm_J = \prod_{j=1}^J \Gm(m_j, n_j, \beta_j),
\end{align}
and the effective Givens rotation angle is given by $\theta_j = \beta_j - \alpha_j$. Substituting \eqref{eq:ujvj} into \eqref{eq:trace_min} and applying the cyclic property of the trace operator, the objective function in \eqref{eq:trace_min} can be expressed as $\mathrm{tr}(\Um_J\tr\bs\Gamma\Vm_J)$.
We then adopt an iterative strategy to place the Givens rotations sequentially, starting with $j = 1$ and proceeding to $ j = J$.  The problem of selecting the rotation for $j=1$ is given by
\begin{align}
    \label{eq:trace_greedy_i}
    \underset{m_1, n_1, \alpha_1, \beta_1}{\max} \;\; &\mathrm{tr}(\Gm(m_1, n_1, -\alpha_1)\bs\Gamma \Gm(m_1, n_1, \beta_1)).
\end{align}
Since the Givens rotation affects only the corresponding $2 \times 2$ submatrix of $\bs{\Gamma}$ indexed by $(m_1, n_1)$, the problem reduces to 
\begin{equation}
    \label{eq:svds}
    \underset{m_1, n_1, \alpha_1, \beta_1}{\max} \;\; \mathrm{tr} \left(\Rm(-\alpha_1)
    \begin{bmatrix}
        a_{m_1m_1} & a_{m_1n_1} \\ a_{n_1m_1} & a_{n_1n_1}
    \end{bmatrix}
    \Rm(\beta_1) \right),
\end{equation}
where $a_{pq}=[\bs{\Gamma}]_{pq}$, and $\Rm(\theta)$ denotes a $2 \times 2$ rotation matrix. 
This problem is combinatorial in nature and NP-hard to solve~\cite{golub2012matrix}. To obtain an approximate solution, we decompose it into two steps: (i) determining the index pair $(m_1, n_1)$, and (ii) computing the corresponding rotation angles $\alpha_1$ and $\beta_1$.

The first problem corresponds to a combinatorial search for a $2 \times 2$ submatrix with the largest sum of singular values~\cite{golub2012matrix}, and this search is computationally expensive. We avoid exhaustively searching over all combinations of index pairs by selecting the rotation location corresponding to the largest absolute off-diagonal entry of the symmetrized right covariance matrix, given by
\begin{equation}
    \label{eq:mn_criteria}
    (m_1, n_1) = \arg\:\underset{(p,q), \: p>q}{\max} \: |[\bs\Gamma\tr\bs\Gamma]_{pq}|.
\end{equation}
After determining the index pair $(m_1, n_1)$, the rotation matrices $\Rm(\alpha_1)$ and $\Rm(\beta_1)$ are obtained from the left and right singular vectors of the $2 \times 2$ matrix in \eqref{eq:svds} by solving a $2 \times 2$ SVD problem~\cite{golub2012matrix}. 
Then, the left and right rotation matrices are updated as $\Um_{1}=\Gm(m_1,n_1,\alpha_1)$,  $\Vm_{1}=\Gm(m_1,n_1,\beta_1)$, and the corresponding transform is given by $\Sm_1 = \Vm_1\Um_1\tr$.
From an RD perspective, our criterion in  \eqref{eq:mn_criteria} identifies and targets the pair of primary transform coefficients with the strongest covariance, and selects an optimal rotation to decorrelate them. Note that using thresholded coefficients $\hat{\Ym}$ in forming $\bs{\Gamma}$ discourages the placement of Givens rotations on coefficient pairs that are likely to be quantized to zero, which aligns the rotation selection process with RD considerations. 
This procedure is repeated iteratively for $j = 2, \ldots, J$. At iteration $j$, we form the rotated cross-covariance matrix $\bs\Gamma_j = \Um_{j-1}\tr\bs\Gamma\Vm_{j-1}$, determine the location $(m_j, n_j)$ such that $(m_j, n_j) \neq (m_i, n_i)$ for all $i<j$, angles $(\alpha_j, \beta_j)$, and update $\Um_{j}=\Um_{j-1}\Gm(m_j,n_j,\alpha_j)$,  $\Vm_{j}=\Vm_{j-1}\Gm(m_j,n_j,\beta_j)$, with the corresponding transform $\Sm_j = \Vm_j \Um_j\tr$. 

\vskip 1ex
\noindent\textbf{Adaptive number of rotations.} The above iterative procedure can be repeated until either the number of rotations reaches a predefined maximum $J_{\max}$ or the remaining off-diagonal energy after applying $J$ rotations, measured by the normalized factorization error~\cite{Magoarou18},
\begin{equation}
    \label{eq:diag_error}
    e_J = \frac{\|[\Um_{J}\tr\bs\Gamma\Vm_{J}]_{\text{off-diag}} \|_F^2}{\|\bs\Gamma\|_F^2}, \quad J = 1, \hdots,n(n-1)/2,
\end{equation}
falls below a predefined threshold $\tau$. Here, $\|\cdot\|_F$ denotes the Frobenius norm, and $[\:\cdot\:]_{\text{off-diag}}$ sets the diagonal entries of a matrix to zero. This stopping criterion provides an adaptive trade-off between decorrelation performance and computational complexity. The complete approximate Givens factorization algorithm used in this work is summarized in \cref{alg:Jacobisvd}. 
\vskip -2ex
\section{Simulations}
\label{sec:empRes}
We use intra-prediction residuals obtained from AV2 for both transform learning and evaluation. Specifically, images from the CLIC dataset~\cite{clic25} are compressed using AV2, and residual blocks of sizes $8 \times 8$, $16 \times 16$, and $32 \times 32$ are collected. Among these blocks, some used the DCT and others the ADST as primary transforms, and some used only a primary transform while others used both primary and secondary transforms. The resulting dataset is split into training and testing sets with a ratio of $4{:}1$, respectively, for each intra-prediction mode. The number of samples used per mode reflects the typical distribution of modes selected by RDO. 

\begin{figure}
    \centering
    \includegraphics[width=1\linewidth]{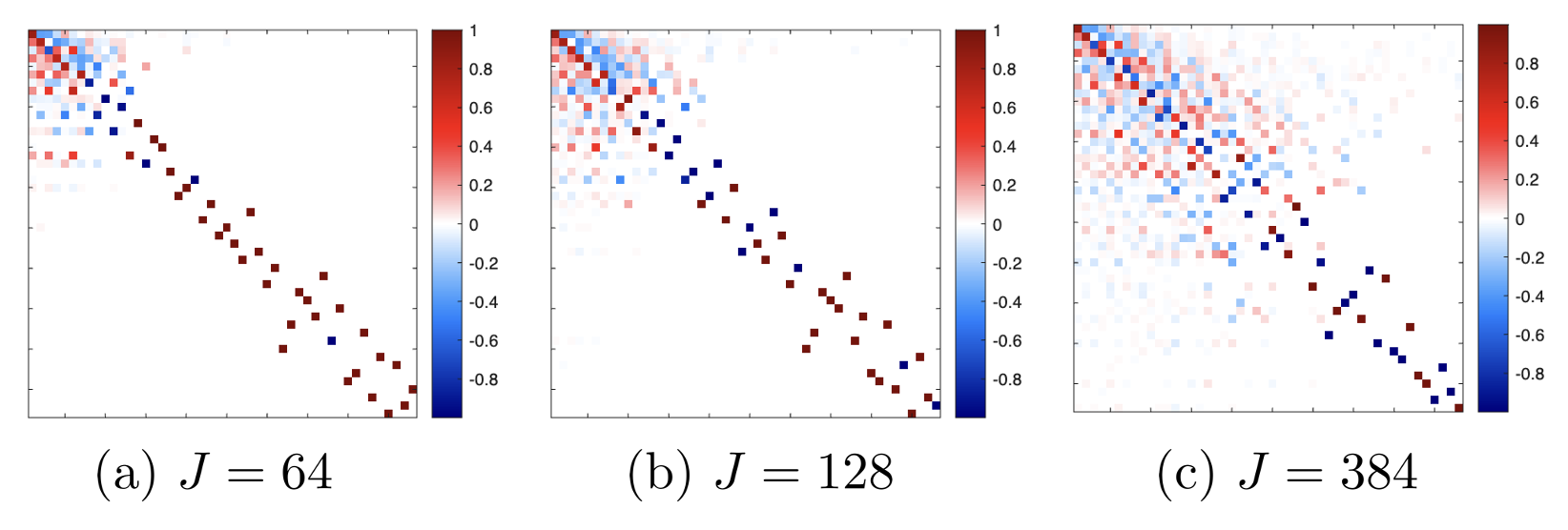}
    \vskip -2ex
    \caption{FaSST kernels designed using $\tau=10^{-6}$ and different numbers of Givens rotations: (a) $J = 64$, (b) $J = 128$, and (c) $J = 384$.}
    \label{fig:FaSST}
    \vskip -2ex
\end{figure}
\noindent\textbf{Mode-dependent secondary transform design.} We design data-dependent secondary transforms for the 12 principal intra-prediction modes in AV2 using a mode-dependent framework. In this setup, each intra-prediction mode can select from four transform options: DCT, ADST, a secondary transform applied after DCT, and a secondary transform applied after ADST. For each mode, a dedicated scanning order is employed to sort the primary transform coefficients according to their expected variance prior to applying the secondary transform. For secondary transform, we consider: (1) KLT, (2) LFNST~\cite{Koo19}, (3) SOT, (4) LF-SOT (SOT retaining only $n_k$ coefficients), (5) KLT-GR (approximating KLT with Givens rotations as a proxy for~\cite{Said16high}), (6) the proposed FaSST with fixed $J$ for all modes, and (7) FaSST with mode-adaptive $J$. Except for LF-NST and LF-SOT, the other methods do not truncate primary or secondary transform coefficients. For each intra-prediction mode, an RDO clustering is performed in the presence of all candidate transforms: DCT, ADST, and one secondary transform following each of DCT and ADST~\cite{Pak25}. 
For KLT-GR and the two variants of FaSST, we use the cluster assignments obtained from KLT- and SOT-based clustering, respectively, and learn the corresponding low-complexity transforms. All transforms are designed separately for block sizes of $8 \times 8$, $16 \times 16$, and $32 \times 32$. 

We let $QP \in \{26, 27, 28, 29, 30, 31\}$ for both training and testing, which correspond to PSNR values in the range of approximately $28$–$40$ dB~\cite{Zhou04psnr}. The quantization step size is $Q_s = 2^{(QP-4)/6}$, and the Lagrange multiplier for RDO is set to $\lambda = 0.85 \times 2^{(QP-12)/3}$ \cite{ringis2023disparity}. For both SOT and FaSST, the sparsity regularization parameter is set to $\mu = (Q_s/2)^2$. To learn a single transform kernel that is effective across the $QP$ range, we adopt an annealing-like training procedure~\cite{Sezer15sot}, starting from the largest $\mu$ value and progressively decreasing it, while using the learned kernels and clustering assignments at each stage to initialize the subsequent optimization. For all methods, the number of primary transform coefficients selected for secondary transform is set to $n = 48$, and for LFNST and LF-SOT, $n_k = 32$ coefficients are retained for all block sizes, consistent with the secondary transform pipeline in AV2. 
For the proposed FaSST framework with fixed $J$, we design transforms using $J_{\max} \in \{64, 128, 192, 256, 384, 512\}$ Givens rotations and set $\tau = 10^{-6}$, which corresponds to approximately $\{11.1\%, 22.2\%, 33.3\%, 44.4\%, 66.7\%, 88.9\%\}$ of the computational cost of the full KLT. For the adaptive-$J$ setting, we fix $J_{\max} = 512$ and vary  $\tau \in \{0.07, 0.05, 0.04, 0.03, 0.02, 0.01\}$ to control the number of selected rotations. All training is carried out offline, and the final transform kernels are made available at both the encoder and decoder.

\begin{figure}[t]
    \centering
    \includegraphics[width=0.65\linewidth]{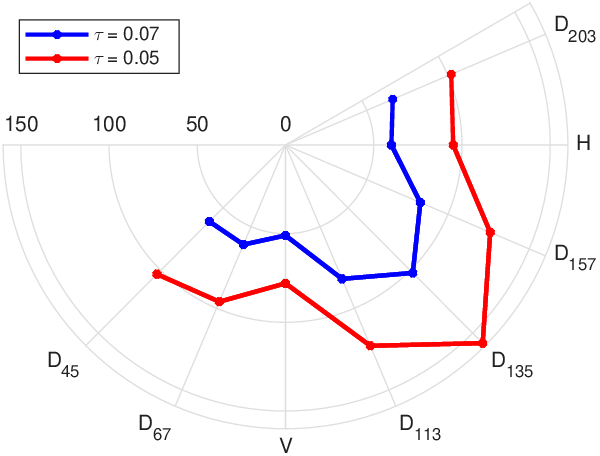}
    \vskip -2ex
    \caption{Number of rotations selected by the mode-adaptive FaSST algorithm as a function of intra-prediction angle and the threshold $\tau$.}
    \label{fig:numrot}
    \vskip -4ex
\end{figure}

We first present the FaSST kernels designed for different values of $J$ in \cref{fig:FaSST} for $16 \times 16$ residual blocks corresponding to the $D_{135}$ prediction mode. When $J$ is small, most Givens rotations are concentrated among low-frequency components, with only a few rotations applied to adjacent higher-frequency components. As $J$ increases, a larger number of rotations are allocated to low- and mid-frequency components, while relatively few rotations are placed at high frequencies. This behavior indicates that the proposed rotation placement strategy effectively accounts for the energy distribution and targets the most significant residual correlations, thereby achieving improved rate–distortion performance under a predefined complexity budget determined by the number of rotations $J$. Next, we illustrate how the adaptive number of rotations is selected for angular intra-prediction modes as a function of the intra-prediction angle in \cref{fig:numrot}. It can be observed that fewer rotations are selected for horizontal and vertical modes, since separable primary transforms leave relatively little residual correlation for these directions. In contrast, a larger number of rotations is allocated for modes such as $D_{135}$, where the directional statistics are more difficult to decorrelate using separable primary transforms alone.

\begin{figure}[t]
    \centering
    \includegraphics[width= 1\linewidth]{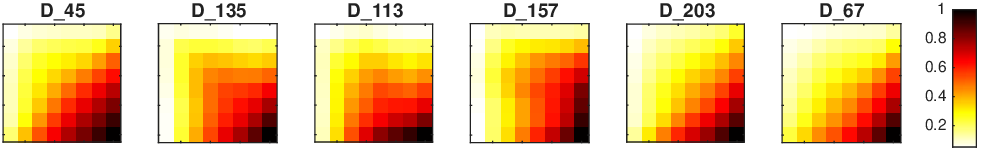}
    \vskip 1.8ex
    \includegraphics[width= 1\linewidth]{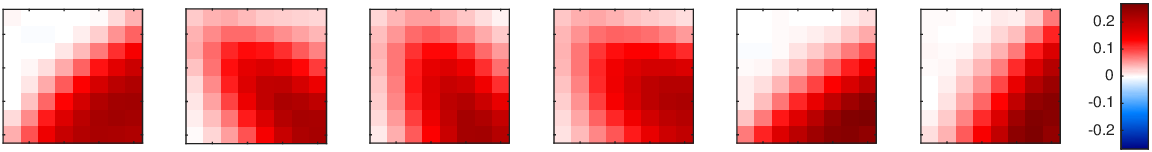}
    \vskip -2ex
    \caption{Sample variance and DC bases of mode-dependent transforms for $8 \times 8$ residuals: (top row) normalized sample variance of the prediction residuals, (bottom row) DC basis (first column of $\Fm$ reshaped into $8 \times 8$ blocks) of the transform $\Fm = \Gm \Pm \Tm$, where $\Tm$ is the FaSST with adaptive $J$ corresponding to $\tau = 0.05$.}
    \label{fig:trbasis}
\end{figure}

Furthermore, we show the lowest frequency basis vectors of the effective transform applied to $8 \times 8$ blocks in \cref{fig:trbasis}, using FaSST with adaptive $J$ followed by DCT for few intra-prediction modes of AV2, along with the normalized sample variance of the residuals for $\tau = 0.05$. We observe that the low-frequency basis vectors adapt to the directional statistics expected for each mode while requiring significantly fewer computations compared to KLT. 

\begin{figure}[t]
    \centering
    \includegraphics[width=0.95\linewidth]{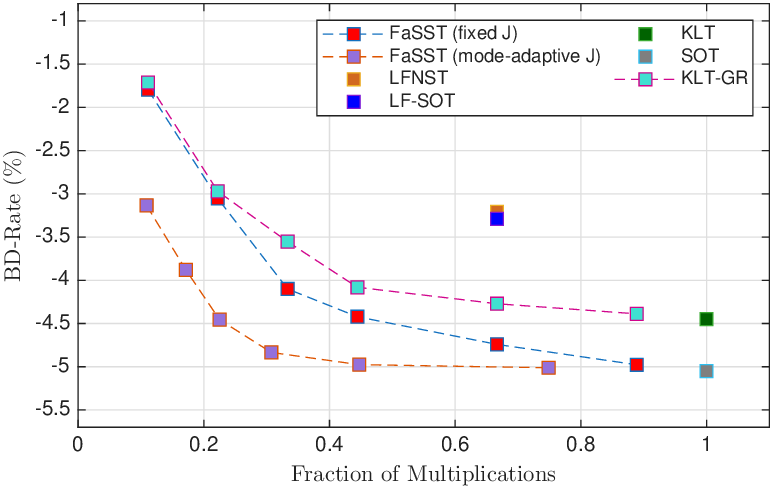}
    \vskip -2ex
    \caption{Average BD-rate savings (more negative is better) versus fraction of multiplications relative to KLT for FaSST (fixed/mode-adaptive $J$), LFNST~\cite{Koo19}, LF-SOT, KLT, SOT~\cite{Sezer15sot}, and KLT-GR~\cite{Said16high}.}.
    \label{fig:results}
    \vskip -5ex
\end{figure}

\noindent\textbf{Evaluation of coding performance.} Next, we show that FaSST offers a better performance–complexity trade-off than existing methods. We independently encode each residual block from the test set. The process involves applying the transform, uniform quantization, and entropy coding using a CABAC encoder~\cite{sze2012cabac}. We first select the best primary transform between DCT and ADST for each block using a standard RDO process. Then, we perform a secondary RDO step to decide whether or not to apply a secondary transform. We use $2$ bits to signal the selected transform for each block: one bit to indicate the primary transform and one to indicate whether a secondary transform is applied. This RDO approach for selecting the best combination of primary and secondary transforms is consistent with state-of-the-art codecs that use secondary transforms~\cite{zhao2021study}.

We compare different secondary transform design methods against a baseline that uses only two primary transforms, DCT and ADST, with quantization, entropy coding, and RDO performed as described. The baseline uses a single bit of overhead to signal the selected transform. For each method, including the baseline, we compute the bitrate and distortion for each block with the same $QP$s used in the training and average the results across block sizes of $8 \times 8$, $16 \times 16$, and $32 \times 32$ for all intra-prediction modes, and report the average Bjøntegaard rate (BD-rate) savings~\cite{bjontegaard2001calculation} relative to the baseline. We use the number of multiplications required to compute the secondary transforms as a measure of computational cost, expressing this cost as a fraction of the number of multiplications required by KLT (i.e., $n \times n$). In the case of FaSST, the total number of multiplications required is $4J$, since each Givens rotation requires 4 multiplications. For the mode-adaptive FaSST, the total number of multiplications across all intra-prediction modes is averaged over the number of modes. 
Moreover, the number of additions required by the KLT is $n(n-1)$, by LFNST is $n_k(n-1)$, whereas FaSST requires $2J$ additions. Since all methods use the same number of kernels, memory requirements scale linearly with the number of multiplications.

The average BD-rate savings versus the fraction of multiplications are shown in \cref{fig:results}. We observe that BD-rate performance comparable to LFNST can be achieved using approximately $33.33\%$ of the computational cost of LFNST with KLT-GR and FaSST configured with a fixed number of rotations. In addition, by adaptively placing rotations with the mode-adaptive FaSST, the computational cost can be further reduced to approximately $16.33\%$ relative to LFNST. Moreover, using only $33.76\%$ of the computational cost of LFNST, the proposed FaSST with a mode-adaptive number of Givens rotations achieves an additional BD-rate saving of $1.80\%$ compared to LFNST. These results demonstrate that FaSST is a promising alternative for reducing the computational complexity of secondary transforms in next-generation video coding systems. Furthermore, the secondary transform pipeline in VVC, including coefficient dropping, is similar to that of AV2 \cite{bross2021overview}. Therefore, our observations, including that coefficient dropping may not be optimal, are expected to generalize to VVC. Experimental validation on VVC is left for future work.

\vskip -2ex
\section{Conclusion}
\vskip -1ex
In this work, we proposed an FaSST framework for designing fast, data-dependent secondary transforms for residual coding. We formulate secondary transform learning as a rate-distortion-cost-inspired SOT problem with the additional constraint that the transform be factorized into a sequence of Givens rotations, enabling efficient implementations. The alternating minimization scheme, combined with a cross-covariance–driven Givens factorization, allows FaSST to selectively target decorrelation of the most significant covariances in the data within a predefined complexity budget. Experimental results on AV2 intra-prediction residuals demonstrate that FaSST achieves RD performance comparable to state-of-the-art LFNST while reducing computational complexity by up to $83.67\%$, and further enables BD-rate savings of $1.80\%$ at only $33.76\%$ of the computational cost of LFNST through mode-adaptive rotation placement. These results indicate that FaSST provides an effective and practical alternative for secondary transform design in modern video codecs. Evaluation of the proposed transforms within a full-reference software implementation is left for future work.

\vskip 1ex
\noindent\textbf{Acknowledgment.}
We thank Keng-Shih Lu and Kruthika Koratti Sivakumar for providing the training datasets and feedback.

\bibliographystyle{IEEEbib}
\bibliography{GSP,VideoCoding}

\end{document}